\documentclass{sigchi-ext}
\usepackage[T1]{fontenc}
\usepackage{textcomp}
\usepackage[scaled=.92]{helvet} 
\usepackage{graphicx} 
\usepackage{balance}  
\usepackage{booktabs} 
\usepackage{ccicons}  
\usepackage{ragged2e} 
\usepackage{todonotes}

\def\plaintitle{SIGCHI Extended Abstracts Sample File: Note Initial
  Caps} 
\def\emptyauthor{}
\def\plainkeywords{Reconnection; Older adults; Social networks; Design challenges; Qualitative study}

\title{Design Challenges for Reconnecting in Later Life: A Qualitative Study}

\numberofauthors{7}
 \author{%
   \alignauthor{%
     \textbf{Francisco Ibarra}\\
     \affaddr{University of Trento} \\
     \affaddr{Trento, Italy} \\
     \email{fj.ibarracaceres@unitn.it} }\alignauthor{%
     \textbf{Norma Lau}\\
     \affaddr{University of Costa Rica}\\
     \affaddr{San Jos\'e, Costa Rica}\\
     \email{norma.lau@ucr.ac.cr} } \vfil \alignauthor{%
     \textbf{Grzegorz Kowalik}\\
     \affaddr{Polish-Japanese Academy}\\
 	\affaddr{of Information Technology}\\
     \affaddr{Warsaw, Poland} \\
     \email{grzegorz.kowalik@pja.edu.pl} }\alignauthor{%
     \textbf{Luca Cernuzzi}\\
     \affaddr{Catholic University}\\
     \affaddr{``Nuestra Seora de la Asunci\'on''}\\
     \affaddr{Asunci\'on, Paraguay}\\
     \email{lcernuzz@uc.edu.py} } \vfil \alignauthor{%
     \textbf{Marcos Baez}\\   
     \affaddr{University of Trento}\\    
     \affaddr{Trento, Italy}\\
   	\affaddr{Tomsk Politechnic University}\\
     \email{baez@disi.unitn.it} }\alignauthor{%
     \textbf{Fabio Casati}\\
     \affaddr{University of Trento}\\
     \affaddr{Trento, Italy}\\
 	\affaddr{Tomsk Politechnic University}\\    
     \email{fabio.casati@unitn.it} } \vfil
 \alignauthor{%
     \textbf{Rados\l{}aw Nielek}\\    
     \affaddr{Polish-Japanese Academy}\\
 	\affaddr{of Information Technology}\\
     \affaddr{Warsaw, Poland} \\
     \email{nielek@pja.edu.pl} 
 }\alignauthor{%
 } }

\definecolor{linkColor}{RGB}{6,125,233}
\hypersetup{%
  pdftitle={\plaintitle},
  pdfauthor={\emptyauthor},
  pdfkeywords={\plainkeywords},
  bookmarksnumbered,
  pdfstartview={FitH},
  colorlinks,
  citecolor=black,
  filecolor=black,
  linkcolor=black,
  urlcolor=linkColor,
  breaklinks=true,
}

\begin{document}

\CopyrightYear{2018} 
\setcopyright{rightsretained} 
\conferenceinfo{DIS'18 Companion}{June 9--13, 2018, , Hong Kong}
\isbn{978-1-4503-5631-2/18/06}
\doi{https://doi.org/10.1145/3197391.3205426}
\copyrightinfo{\acmcopyright}

\maketitle

\RaggedRight{} 

\begin{abstract}
  Friendships and social interactions are renown contributors to wellbeing. As such, keeping a healthy amount of relationships becomes very important as people age and the size of their social network tends to decrease. 
In this paper, we take a step back and explore reconnection \textemdash find out about or re-contact old friends, an emerging topic due to the increased use of computer-mediated technology by older adults to maintain friendships and form new ones.
We report on our findings from semi-structured interviews with 28 individuals from Costa Rica and Poland. The interviews aimed to explore whether there is a wish to reconnect, and the challenges encountered by older adults to reconnect. 
We contribute with design considerations for tools allowing older adults to reconnect, discussing opportunities for technology.
\end{abstract}


\category{H.5.m.}{Information Interfaces and Presentation
  (e.g. HCI)}{Miscellaneous} {}{}

\section{Introduction}

Friends and family play an essential role in our lives, providing instrumental and emotional support that contributes to our general health and wellbeing. This is true at every age and even more as we grow older \cite{holt2015loneliness}. 
Across the life span, this network of support tends to become smaller \cite{wrzus2013social} and more geographically dispersed \cite{ajrouch2005social}. 
These changes, along with life events such as retirement or bereavement, limit our opportunities to engage in social interactions, thus putting us at risk of loneliness and social isolation \cite{pinquart2001influences,havens2004social}.

\begin{marginfigure}[1pc]
\vspace{-5pt}
  \begin{minipage}{.97\marginparwidth}
    \centering
  \includegraphics[width=\columnwidth]{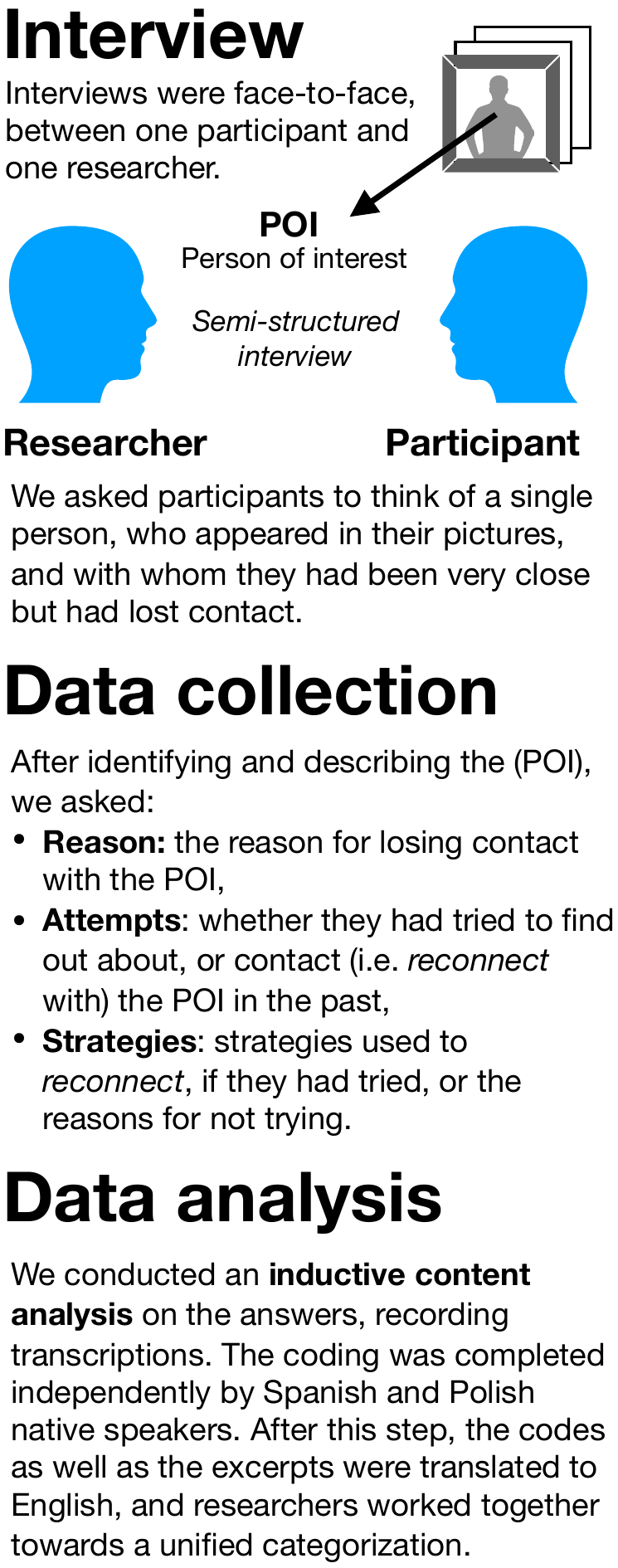}
  \captionof{figure}{Study procedure}
  \label{fig:fig1}
  \end{minipage}  
\end{marginfigure}

\begin{margintable}[1pc]
  \begin{minipage}{\marginparwidth}
    \centering
    \begin{tabular}{r l}
     \textbf{Location} (n) & \textbf{Age} (Mean)  \\
      \toprule
      Costa R. (17) & 66-90 (70.5) \\
      \midrule
      Poland (11) & 67-82 (70) \\
      \bottomrule
    \end{tabular}
    \caption{Study participants.}~\label{tab:table1}
  \end{minipage}
\end{margintable}

Research on social interactions suggests distinct benefits and challenges of familial and friendship relationships. 
Friendships are associated with stronger effects on subjective wellbeing \cite{pinquart2000influences, mellor2008can}, impact on morale \cite{wood1978friendship} and better functioning \cite{chopik2017associations} as compared to familial relationships. 
This relative importance of friendships in later life is explained by the voluntary nature of friendships, which makes them more selective and therefore, potentially of higher quality \cite{wood1978friendship,adams2015friendship}. 
For the same reasons, friendships are also more vulnerable than any other type of relation \cite{adams1986look}, especially to non-normative life events such as relocation \cite{wrzus2013social}. 
Indeed, the reduction in the social network size with age is particularly present in the friendship social network \cite{van1998losing,wrzus2013social}.

In this paper we aim at understanding the challenges and opportunities in \emph{reconnecting} older adults with lost friends and contacts with the help of technology. 
Reconnecting is an emerging topic \cite{jung2016senior,jung2017social,desiato2013virtualizing} motivated by older adults' increasing use of computer-mediated technology, particularly social media, to maintain and re-engage social connections.

Surveys with older adult users of Facebook, for instance, indicate usage to be driven mainly by the need to support and maintain existing social networks, by \emph{checking} what friends are up to, \emph{keeping} in touch with friends, and \emph{re-acquiring} lost contacts \cite{jung2016senior,jung2017social}. The latter is cited not only a source of great user satisfaction but as an opportunity to build social capital \cite{joinson2008looking}. However, our understanding of reconnection is limited to studies on specific social media platforms and by older adults from specific regions.
Such limitations motivate a better understanding of how to support reconnection, and what existing or new technologies may support reconnection efforts.

In this paper, we report on a pilot for a larger multi-site study aiming at exploring the potential of reconnection in later life. We focus on the fundamental question of whether there is a wish for reconnecting, and the reasons why older adults fail to reconnect with their friends. In doing so, we aim to identify the main challenges that can help derive design considerations for technology as well as guide future reconnection studies.

The pilot was conducted in Costa Rica and Poland, locations representing different social and cultural contexts. 
We interviewed 28 participants and found that the wish for reconnecting is real and that the challenges older adults face are only partially addressed by current technology. In the remainder we describe the pilot and our preliminary results.

\section{Methods}

In Costa Rica, we studied a rural community near the capital, characterized by very low income, very high crime rate, and high social ties within the community. In Poland, we conducted the study in a small city in a rural area with students of the University of the Third Age, where retired or semi-retired older adults can continue their education and training. 
In selecting these two locations, our aim was to test the pilot under different conditions, and assess whether the different social, cultural and economic contexts had an impact on the wish and challenges to reconnect.

\newpage
We considered eligible, participants who had no cognitive or vision impairments and were 65 or older (age to be considered an older person in developed countries \cite{who2002}).
We asked participants for pictures related to school, work, vacations, trips, or events (e.g. concerts), dating back at least 10 years and portraying three or more (non-relatives) people. 
These pictures would later be used to identify people with whom they were very close but had lost contact, people of interest (POI).
The specific objective of reconnecting was not disclosed, to prevent participants from thinking of people with whom it would be easier to reconnect.

In Costa Rica participants were recruited via phone calls, from a directory of people who already agreed to participate to a social project organized by the University of Costa Rica, of which this study is part. In Poland, recruitment was done face-to-face at the University of the Third Age.

\begin{marginfigure}[0pc]
\vspace{-5pt}
  \begin{minipage}{.97\marginparwidth}
    \centering
  \includegraphics[width=\columnwidth]{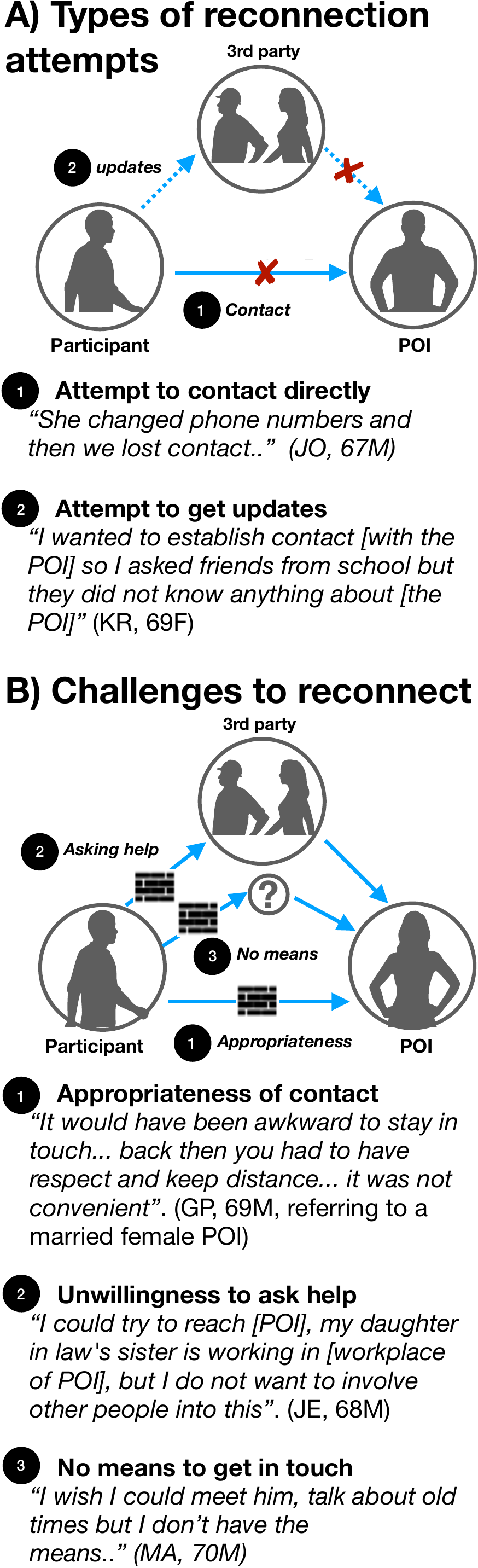}
  \captionof{figure}{Reconnection attempts and challenges}
  \label{fig:fig2}
  \end{minipage}  
\end{marginfigure}

We conducted semi-structured interviews with 38 older adults (26 female and 12 male). We discarded 10 interviews from Costa Rica from the analysis, for a total of 28 participants (18 female and 10 male, age range: 66-90, mean: 71.8). Interviews were omitted for the following (non-exclusive) reasons: 
participant had recent contact with POI (7), participant was related to POI (2), participant did not indicate a POI (3), and POI had passed away (1).
In the interview, participants were asked to identify a POI from the pictures, and inquired on their wish and attempts made to reconnect. The data collected was analyzed using inductive content analysis. Details are shown in Figure \ref{fig:fig1}. The study was approved by the IRB of the University of Costa Rica.

\section{Findings}

\subsection{The wish to Reconnect}
Our results show that the problem of "losing contacts" is real: We identified friends with whom people had lost contact in nearly all interviews (barring 3 participants who could not identify a POI among the pictures they brought). In most cases these were friends from school, work, or neighbors. Moreover, only for 4 out of 28 lost contacts from \emph{randomly} selected pictures there was no interest in reconnecting. 

The main reasons for disconnecting were people relocating or the association ending (e.g., people changing jobs, finishing school), thereby ceasing the occasion for contact. For relocation, however, economic and social factors differ (e.g., migration and employment instability in Costa Rica and moving to study in Poland). These observations support current literature on the vulnerability of friendship relationships \cite{adams1986look,ajrouch2005social} and on how life events shape our social networks making them smaller and more dispersed \cite{wrzus2013social,ajrouch2005social}. 

The wish to reconnect was manifest in two distinctive ways (see Figure~\ref{fig:fig2}A): \emph{resuming contact} with the lost friend, e.g., by trying to contact them directly, and \emph{checking out} what the lost friend is up to without necessarily engaging in a conversation, e.g., asking a friend about them. These two scenarios are closely related to the motivations of older adults who actually use online social networks \cite{jung2016senior,jung2017social}.

\subsection{Challenges to reconnection}
Practical and perceived barriers prevented participants to attempt or realize their wish to reconnect. 
In most cases, the interest in reconnecting was not followed by actual attempts (17 participants did not try, out of which 4 expressed no interest in reconnecting). 
Only 5 participants acted (unsuccessfully) on their wish, while other 6 tried only to get updates about their lost friends through common contacts.
Those who did not try to reconnect, mainly did not because they i) did not know how or have the means to get in touch (9 people had no way to contact or had lost track POIs), and ii) felt awkward just "popping up" in people's life after a long time (see Figure~\ref{fig:fig2}B).

Digging deeper on the awkwardness aspect, we identified two salient aspects. The first was a certain discomfort in initiating a call to reconnect. Another issue was the "appropriateness" of contact (e.g., contacting a now married female friend).
Thus, the challenge might not only be finding a way to initiate contact but also creating opportunities comfortable for both sides to write or talk. This observation is supported by a recent study on factors affecting users decisions to accept friend requests on Facebook from people who had been out of touch for long periods, indicating the influence of factors such as social anxiety, sociability, uncertainty about the partner and perceived reward \cite{ramirez2016reconnect}.

\begin{marginfigure}[.5pc]
  \begin{minipage}{\marginparwidth}
    \centering
  \includegraphics[width=\columnwidth]{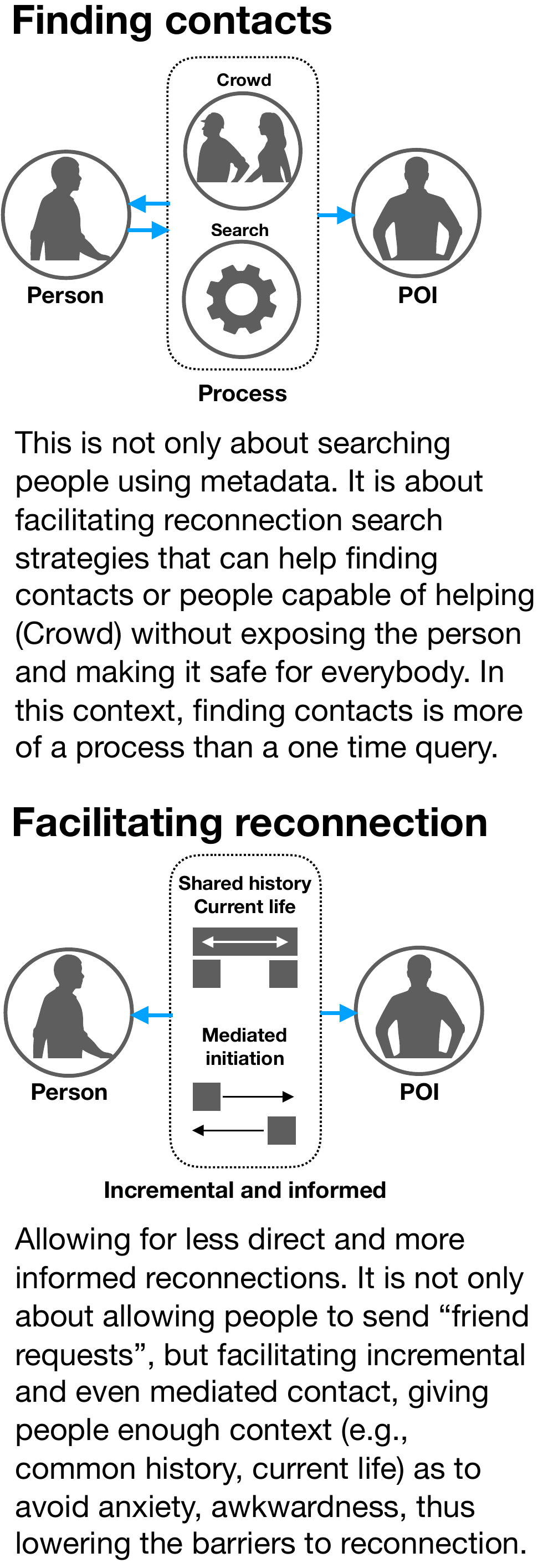}
  \captionof{figure}{Opportunities for technology}
  \label{fig:fig3}
  \end{minipage}  
\end{marginfigure}

\subsubsection{Opportunities for technology}

What we take home here as a first observation is the wish for a "third party" to be involved in creating opportunities for reconnection that takes away the burden of initiating contact and the related awkwardness (e.g.,facilitating a gradual or casual reconnection). Participants referred to this as persons who would organize a reunion, but we speculate that an IT system could play the same (or a supporting) role. 

This being said, participants who had last met their contacts in reunions reported the interest in keeping in touch fading after some time. It is therefore clear that reconnection and relational maintenance should go hand in hand. 
This hints to the importance of social context and associations \cite{quinn2013we}, and building on common aspects \cite{sanchiz2016makes}, so as to motivate meaningful reconnection and strengthen bonds.
IT systems should support this need by helping older adults keep connections active or reviving dormant ties (as suggested by previous studies, e.g., \cite{lim2013reviving}).

With respect to finding contact information on the lost friend. Interestingly, not having the contact at hand, and so having to ask others for help, sets the barrier to reconnecting too high for some participants. Social search features can lower these barriers by allowing users to find their friends online. Still, the limited personal information shared by older adults on social media \cite{hutto2015social} and the perceived complexity \cite{jung2017social} might pose challenges to an effective use of these tools. 

\section{Discussion and Conclusion}
 
Our preliminary studies point to a wish for reconnecting in later life, independently of the cultural context. Although besides this wish, we observed that only respondents from Poland did not want to involve other people when trying to reconnect. This hints to the importance of cultural aspects in the design of suitable solutions.
Current technology gives an opportunity to do so, since it fails to address some of the barriers that prevent reconnection attempts. Social networks are a good foundation, but we argue
that technology should assist on initiating contact, help to find information and keep (re)connection active and meaningful (see Figure \ref{fig:fig3}). Search should be more about 
facilitating strategies that take advantage of the help our network or others can provide to find lost contacts, friendship requests should provide context to reduce awkwardness and become less intrusive. 

In focusing on failed reconnection cases we have added a new perspective to the emerging reconnection phenomena. Reconnecting is a complex construct that requires further studies on how to facilitate this process. This pilot also helped us identify limitations and considerations for future studies, including measuring the perceived connectedness with POI, explore more in detail the role of technology, and the perceived gain in reconnecting.

\section{Acknowledgements}
This project has received funding from the EU Horizon 2020 research and innovation programme under the Marie Sk\l{}odowska-Curie grant agreement No 690962.

\balance{} 

\bibliographystyle{SIGCHI-Reference-Format}
\bibliography{friends,isolation,reconect,ref}

\end{document}